\documentclass[twocolumn,twoside,dvips,showpacs,aps,prl,amsmath,amssymb]{revtex4}

\usepackage{graphicx}
\usepackage{dcolumn}
\usepackage{bm}
\usepackage{amsmath}
\usepackage{amssymb}
\usepackage{bbm}
\usepackage{mathrsfs}
\usepackage[sans]{dsfont}
\usepackage{epsfig,psfrag}
\usepackage{amsmath,amsfonts,amssymb}
\usepackage[usenames]{color}
\definecolor{darkblue}{rgb}{0,0.1,0.5}
\usepackage[dvips, bookmarks, colorlinks=true, plainpages = false, citecolor = darkblue, linkcolor = darkblue, urlcolor = darkblue, filecolor = darkblue]{hyperref}
\def\be{\begin{equation}}
\def\ee{\end{equation}}
\def\bea{\begin{eqnarray}}
\def\eea{\end{eqnarray}}

\begin{document}

\title{Percolation description of the global topography of Earth and Moon}

\author{ Abbas Ali Saberi }\email{ab.saberi@ut.ac.ir}

\address {Department of Physics, College of Science,  University of Tehran, Post
  Office Box 14395-547, Tehran, Iran \\ Institut f\"ur Theoretische
  Physik, Universit\"at zu K\"oln, Z\"ulpicher Str. 77, 50937 K\"oln,
  Germany}

\date{\today}

\begin{abstract}
Remarkable global correlations exist between geometrical features of
terrestrial surface on the Earth, current mean sea level and its
geological internal processes whose origins have remained an
essential goal in the Earth sciences. Theoretical modeling of the
ubiquitous self-similar fractal patterns observed on the Earth and
their underlying rules is indeed of great importance. Here I present
a percolation description of the global topography of the Earth in
which the present mean sea level is automatically singled out as a
critical level in the model. This finding elucidates the origins of
the appearance of scale invariant patterns on the Earth. The
criticality is shown to be accompanied by a continental aggregation,
unraveling an important correlation between the water and long-range
topographic evolutions. To have a comparison point in hand, I apply
such analysis onto the lunar topography which reveals various
characteristic features of the Moon.
\end{abstract}


\maketitle

Discovering the connection between geometrical features of
terrestrial surface on the Earth and its geological internal
processes has long been a basic challenge area in the Earth sciences
\cite{Lyell, Cazenave}, and attracted the attention of physicists
and mathematicians as well \cite{Perrin}. Various theoretical models
have emerged to identify the underlying constructive rules
responsible for the appearance of self-similarity, scale and
conformal invariance in the fractal geometry of the local geomorphic
patterns \cite{Mandelbrot1, Mandelbrot2, Gagnon1, Gagnon2,
Baldassarri, Boffetta}. In comparison with different models proposed
to describe the statistical properties of regional features
\cite{Stark, Maritan, Sapoval}, the global topography has received
less attention and thus remained controversial. Here I show that the
global surface topography can be well described by percolation
theory, the simplest and fundamental model in statistical mechanics
that exhibits phase transitions. A dynamic \emph{geoid}-like level
is defined as an equipotential spherical surface as a counterpart of
the percolation parameter. The analysis shows a geometrical phase
transition in which the critical level surface directly corresponds
to the present mean sea level on the Earth, automatically singling
this level out. This may shed new light on the tectonic plate motion
and help unravel the dynamic story of the Earth's interior. As a
comparison, I also present its application to the lunar topography.

\begin{figure}[b]
  \[
  \includegraphics[width=0.43\textwidth]{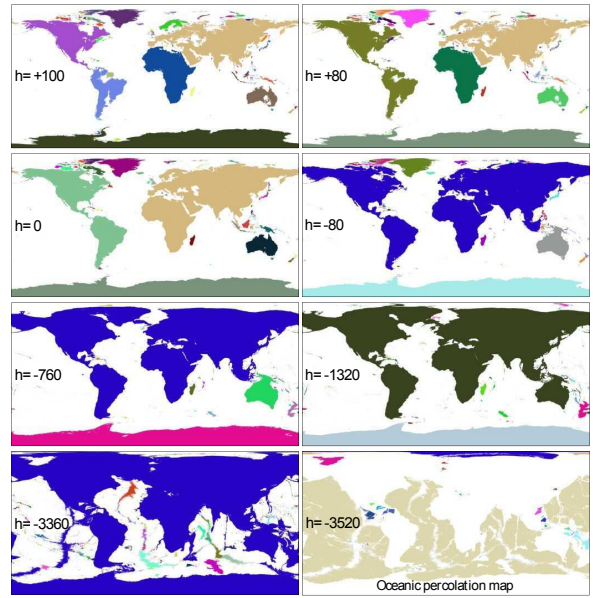}
  \]
  \caption{(Color online) Schematic illustration of the continental aggregation by decreasing the sea level from top to bottom. The first four snapshots
  are selected around the remarkable percolation transition at the present mean sea level around which the major parts of the landmass join together.
  This is followed by three other levels indicating the junction of the Greenland, Australia and Antarctica to the giant landmass.
  The lowest right figure shows the longitudinal percolation of the oceanic clusters (see the supplementary material for enlarged figures
  and also the ones at additional sea levels).
    \label{Fig0}}
\end{figure}

Scale invariance is a remarkable feature of the Earth surface
topography. Observations \cite{Vening, Mandelbrot3, Thomas,
Pelletier} indicate that, over a wide range of scales, the power
spectrum $S$ of linear transects of the Earth's topography follows
the scaling relation $S\propto k^{-2}$ with the wave number $k$.
Such a power law spectrum in the topography leads to the
corresponding iso-height lines (such as coastlines) being fractal
sets characterized by a dominant fractal dimension of 4/3
\cite{Perrin,Mandelbrot1,Steinhaus}. Many other ubiquitous scaling
relations observed in the various terrestrial features e.g., in the
radiation fields of volcanoes \cite{Harvey,Gaonac}, surface magnetic
susceptibility \cite{Pilkington}, geomagnetism \cite{Pecknold} and
surface hydrology such as in the river basin geomorphology
\cite{Rodriguez} are all relevant to the wide range scale invariance
of the topography. Nevertheless, further surveys based on fractional
Brownian motion (fBm) model \cite{Mandelbrot3} of
topography/bathymetry \footnote{bathymetry is the underwater
equivalent to topography.} revealed a more complex multifractal
structure of the Earth's morphology giving rise to distinct scaling
properties of oceans, continents and continental margins
\cite{Gagnon2}. Such a difference is also evident in the well-known
bimodal distribution of the Earth's topography \cite{Wegener} that
reflects the topographic dichotomy of continents and ocean basins, a
consequence of plate tectonic processes.

\begin{figure}
  \[
  \includegraphics[width=0.35\textwidth]{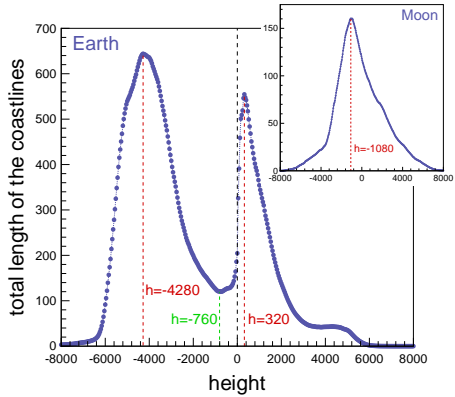}
  \]
  \caption{(Color online) The total length of the coastlines as a function of the sea level (or height)
  for the Earth (main panel) and the Moon (inset).
    \label{Fig1}}
\end{figure}

Plate tectonic theory provides a framework that explains most of the
major surface topographic features of the Earth. It also accounts
for the connection between the processes that facilitate heat loss
and the forces that drive plate motion. The distribution of the
plate areas covering the Earth has been shown to be a power law with
exponent $\sim0.25$ for all plates \cite{Bird,Sornette}. A
remarkable relationship that provides one of the cornerstones of
plate tectonics is that, to a very good approximation, the depth of
the ocean floor beneath the ridge crest increases with square root
of the age of the ocean floor, at least for ocean lithosphere
younger than about $80$ Myr ago \cite{Turcotte}. It plays an
important role in topographical changes and fundamentally affects
long-term variations in global sea level that would assume a surface
equal to the geoid. Here I present the results of a statistical
analysis based on percolation theory that provides new insight into
the better understanding of various interrelationships between the
above mentioned issues and their origins.

I use the topographic data available for the global relief model of
the Earth's surface that integrates land topography and ocean
bathymetry \cite{ETOPO1} (the data information is presented in
http://www.ngdc.noaa.gov/mgg/global/global.html). Current mean sea
level is assumed as a vertical datum of the height relief which
means that the data considers the imperfect ellipsoidal shape of the
Earth. The height relief $\mathfrak{h}$($r$, $\theta$, $\phi$) is
therefore assumed on a sphere of unit radius $r=1$, that also
coincides with the present mean sea level (as zero height level) on
the Earth. All corresponding lengths here are expressed in units of
the Earth's average radius.\\Now imagine flooding this global
landscape in a way that the continental land masses were
criss-crossed by a series of narrow channels so that the resulting
sea level all over the Earth would coincide with a spherical
surface$-$the geoid. All parts above the water level are then
colored differently as disjoint islands, and the rest is left white
(Fig. \ref{Fig0}). If the water level is high, there will be small
disconnected islands, and if it is low, there will be disconnected
lakes. However, there may be a critical value of the sea level
$\mathfrak{h}=h_c$ at which a percolation transition takes place
\cite{cardy, sahimi}.

The percolation problem \cite{sahimi,SA} is an example of the
simplest pure geometrical phase transitions with nontrivial critical
behavior, and it is closely related to the surface topography
\cite{saberi1,saberi2}. At the critical point in two dimensions, the
percolation clusters are some fractal objects whose outer perimeter
is described by a fractal dimension of $4/3$. By considering the
dynamic sea level (height) as a percolation parameter, I examine a
possible description of Earth's topography by means of the
percolation theory.

\begin{figure}[b]
  \[
  \includegraphics[width=0.4\textwidth]{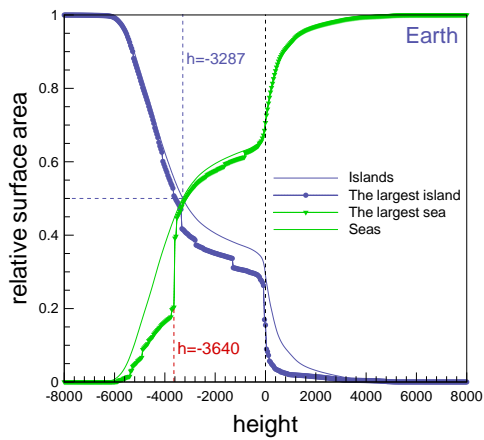}
  \]
  \caption{(Color online) Relative surface area of the largest island (circles) and the largest sea (triangles)
  followed by the total surface area of the islands and the oceans (solid lines) to the total area $4\pi$ of the Earth, as a function of the sea level.
  One critical level is distinguished by each order parameter. The oceanic critical level is close to the level $h=-3287$ m at which
  the total island and oceanic surface areas are equal.
    \label{Fig2}}
\end{figure}

The first quantity of interest is the total length of the coastlines
at varying altitude (or sea level) which is shown in Fig.
\ref{Fig1}. In all figures, the error bars are of the same order as
the symbol size \footnote{The ETOPO1 global relief model has an
accuracy of about $10$ meters at best, likely less accurate in the
deep ocean. Part of the reason for the large error bars are that
each cell's elevation value represents the average elevation over
the entire roughly $2\times2$ km$^2$ footprint of the cell
\cite{ETOPO1}. All mentioned height levels here would indeed bear
such uncertainties.}. Having looked at Fig. \ref{Fig1}, this
quantity closely resembles the height distribution function of the
Earth and the Moon \cite{Stoddard}. The one for the Earth is
characterized by the presence of two levels centered around the
elevations $320$ m and $-4280$ m in the continental platforms and
oceanic floors, respectively. The ratio of the total length of the
coastlines at 320 m to the zero height level is $\sim2.71$. Unlike
Earth, the Moon's curve features only a single peak at around
$-1080$ m.

The usual order parameter is defined as the probability of any site
to be part of the largest island. As shown in Fig. \ref{Fig2}, the
order parameter for islands has a sharp drop-off around the zero
height level i.e, right at the present mean sea level. According to
the further evidence given in the following, it is an indicative of
a geometrical phase transition at this level. The same analysis for
the oceanic clusters (where disjoint oceans at each level are
differently colored, leaving islands white) gives rise to a
discontinuous jump in the oceanic order parameter at around $-3640$
m (Fig. \ref{Fig2}).

\begin{figure}
  \[
  \includegraphics[width=0.4\textwidth]{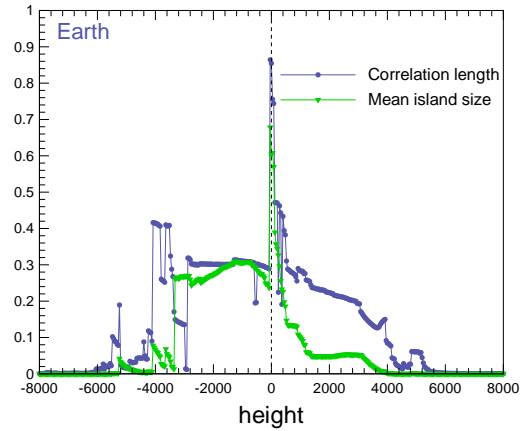}
  \]
  \caption{(Color online) Correlation length and mean island (land) size vs the sea level, with a remarkable characterization of a geometrical
  phase transition at the present mean sea level (zero hight level).
  The scale factors by which all the numbers labeling the vertical axis should be multiplied
  to get the correct graph units are the radius and the square of the radius of the
  Earth for $\xi$ and $\chi$, respectively.
    \label{Fig3}}
\end{figure}

Figure \ref{Fig3} illustrates two other percolation observables
measured for the Earth, the mean island size (analogous to the
susceptibility of the system), and the correlation length. The mean
island size $\chi$ is defined as $\chi=\sum'_s s^2 n_s(h)/\sum'_s s
n_s(h)$, where $n_s(h)$ denotes the average number of islands of
size $s$ at level $h$, and the prime on the sums indicates the
exclusion of the largest island in each measurement. The correlation
length $\xi$ is also defined as average distance of sites belonging
to the same island, $\xi^2=\sum'_s 2R_s^2 s^2 n_s(h)/\sum'_s s^2
n_s(h)$, where $R_s$ is the radius of gyration of a given
$s$-cluster. As shown in Fig. \ref{Fig3}, both quantities $\chi$ and
$\xi$ become divergent at the present mean sea level. The divergence
of the correlation length is a signature of a continuous phase
transition at this level, implying that the critical fluctuations
dominate at each length scale and that the system becomes scale
invariant. These results provide a strong correlation between the
water and long-range topographic evolutions on the Earth.
Nevertheless, one may imagine a model in which water
itself$-$through erosion, evaporation, precipitation and
sedimentation, etc.$-$may have an active role in shaping topography
i.e., the activity of water itself with resulting plains of little
height, shapes the landscape to appear critical around the zero
height.

The measurement of $\chi$ and $\xi$ for the oceanic clusters shows a
dominant divergence that signals the oceanic critical level already
observed in Fig. \ref{Fig2}. The mean ocean size reaches its
absolute maximum at this critical level and the correlation length
remains approximately constant at its maximum for level interval
$-4280\lesssim h\lesssim-3760$ (see the supplementary material).

%

Figure \ref{Fig0} gave an illustration of the percolation transition
at the present mean sea level. As can be seen from the figure, all
major continental junctions occur at the level interval $-80\lesssim
h\lesssim+80$. At a sea level around $-760$ m, Greenland joins the
Afro-Eurasia supercontinent to the Americas and at the same level
the total length of the coastlines reaches its minimum (Fig.
\ref{Fig1}). Australia and Antarctica continents join to the
landmass at $-1320$ m and $-3360$ m, respectively (Fig. \ref{Fig0}).

In order to have a reference point for comparison, as an example of
the most heavily studied waterless body with a completely different
surface properties and interior mechanism, let me analyze the lunar
topography. I used the topogrd2 data set (accessible from
http://pds-geosciences.wustl.edu/missions/clementine/gravtopo.html),
which is measured
relative to a spheroid of radius $1738$ km at the equator$-$the zero height level. I rescale the Moon's average radius to $1$.\\

\begin{figure}[b]
  \[
  \includegraphics[width=0.4\textwidth]{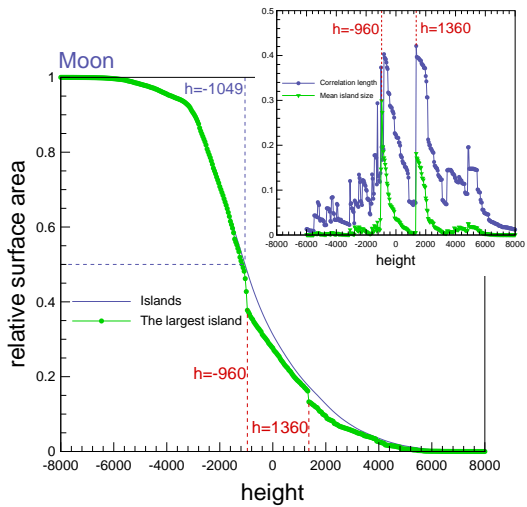}
  \]
  \caption{(Color online) Relative surface area of the largest island followed by the total surface area of the islands
  to the total area $4\pi$ of the Moon, as a function of the hypothetical sea level. At level $\sim-1049$ m,
  the total island and oceanic surface areas on the Moon are equal. The two jumps in the order
  parameter at levels around $-960$ m and $1360$ m are also decoded in the divergent behavior of the correlation length and the
  mean island size (the inset). Such two jumps are unusual for percolation.
    \label{Fig5}}
\end{figure}

\begin{figure}
  \[
  \includegraphics[width=0.45\textwidth]{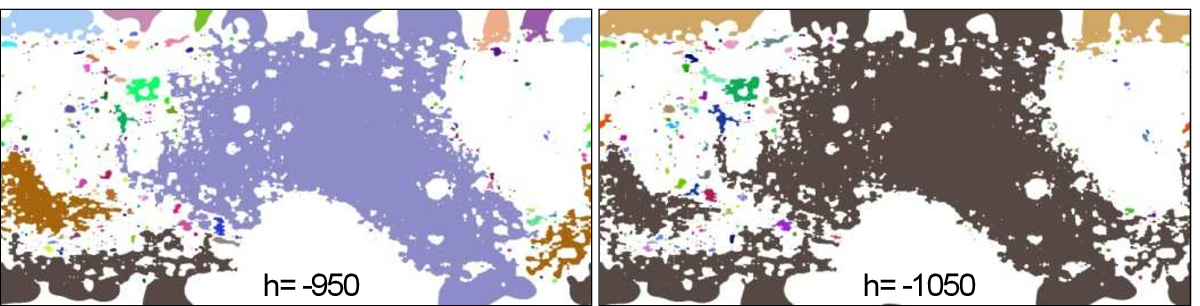}
   \]\[
  \includegraphics[width=0.45\textwidth]{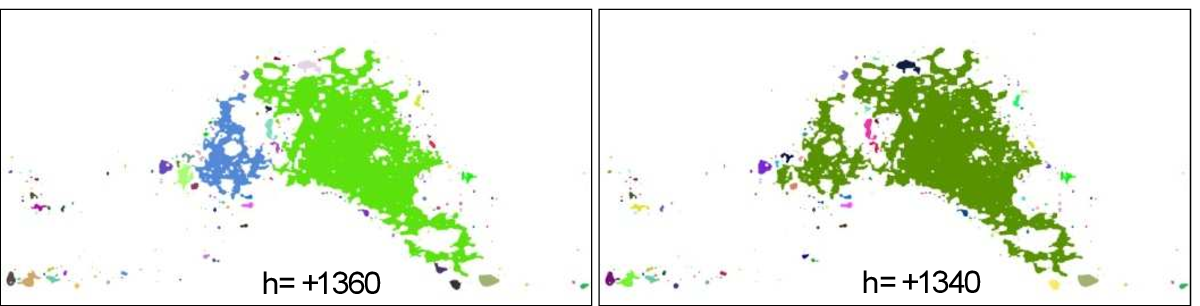}
   \]
  \caption{(Color online) Upper: disjoint islands with heights higher than the hypothetical sea level on the Moon for two different levels $h=-950$ m and $-1050$ m,
  slightly above and below the critical level, respectively. Appearance of the spanning island (right figure) is an indicative
  of a percolation transition. Lower: aggregation of the two main mountainous islands at the lunar farside around the second critical level
  at $\sim1360$ m.
    \label{Fig6}}
\end{figure}

The percolation observables discussed above for the islands are
measured as a function of a hypothetical sea level (Fig.
\ref{Fig5}). The order parameter shows two rather small jumps at
altitude levels around $-960$ m and $1360$ m. The correlation length
and the mean island size have also two dominant peaks at these
levels (see the inset of Fig. \ref{Fig5}).\\The Moon's height
distribution function features a single global peak at level
$\sim-950$ m which is quite close to the one of the critical levels
located at $\sim-960$ m. In addition, if we measure the correlation
length and the mean cluster size for the oceanic clusters, they show
only one critical level very close to the one located at $\sim-960$
m. These may imply that this critical level is more important for
the description of the global topography of the Moon. This is also
quite close to the level $h=-1049$ m at
which the total island and oceanic surface areas are equal, meaning that the island and oceanic percolation thresholds coincide.\\
The illustrative figure \ref{Fig6} shows the connectivity of the
islands on both sides of the critical level. At a height level
$h=-950$ m, a little above the critical level, there exists a number
of disjoint islands. At slightly below the critical level at
$h=-1050$ m, the islands merge together to form a giant percolating
island which spans the Moon in the longitudinal direction.

Nevertheless, the other critical level at $h=1360$ m, unravels a
characteristic feature of the lunar farside. As it is known, one of
the most striking geological features of the Moon is the elevation
dichotomy \cite{Zuber} between the hemispheres: the nearside is low
and flat, dominated by volcanic maria, whereas the farside is
mountainous and deeply cratered. The illustrations in Fig.
\ref{Fig6} for elevation levels at $h=1360$ and $1340$ m, at both
sides of the critical level, indicate the aggregation of two main
mountainous islands that are separated by a very narrow passageway.
This may be a benchmark of a rather non-random origin of the
formation of the lunar farside highlands \cite{Jutzi}.

To summarize, the percolation description of the global Earth's
topography uncovers the important role that is played by the water
on the Earth. The critical threshold of the model coincides with the
current mean sea level on the Earth. This criticality is along with
a sign of the continental aggregation at this level which seems to
be more dominated by the \emph{endogenic} processes (like volcanic
activity, Earthquakes and tectonic processes) originating within the
Earth that are mainly responsible for the very long-wavelength
topography of the Earth's surface, rather than by the
\emph{exogenic} processes like erosion, weathering and
precipitation. The criticality of the current sea level also
justifies the appearance of the scale (and conformal) invariant
features on the Earth with an intriguing coincidence of the dominant
$4/3$ fractal dimension in the critical model and observation. The
main critical level for the Moon has the same amount of land and
oceans at the threshold, indicating a purely geometrical phase
transition.

I would like to thank J. Cardy, M. Kardar, J. Krug, T. Quella and M.
Sahimi for their many useful comments, and especially thank D.
Stauffer for his many helpful comments and suggestions. The author
also thanks H. Dashti-Naserabadi for his kind help with programming
and M.J. Fallahi for the Earth's data source. Financial support from
the Deutsche Forschungsgemeinschaft via SFB/TR 12 and supports from
Alexander von Humboldt Foundation, are gratefully acknowledged. I
also acknowledge partial financial supports by the research council
of the University of Tehran.

\end{document}